# GEODESIC FOLDING OF TETRAHEDRON

SERI NISHIMOTO, TAKASHI HORIYAMA, AND TOMOHIRO TACHI

*Name: Seri Nishimoto (b. Japan)*
*Profession: Student (Architecture)*
*Fields of interest: Geometric Design, Architectural Planning*
*E-mail: s-nishimoto@g.ecc.u-tokyo.ac.jp*

*Name: Takashi Horiyama (b. Japan)*
*Profession: Associate Professor*
*Fields of interest: Computational Geometry, Enumeration Algorithms*
*Address: 255 Shimo-Okubo, Sakura, Saitama 338-8570, Japan*
*E-mail: horiyama@al.ics.saitama-u.ac.jp*
*Home-page: http://www.al.ics.saitama-u.ac.jp/horiyama/*
*Awards: EATCS/LA Best Presentation Award*
*Publications and/or Exhibitions: T. Horiyama, W. Shoji (2011); Edge Unfoldings of Platonic Solids Never Overlap, In Proc. of the 23rd Canadian Conference on Computational Geometry, pp. 65-70.*

*Name: Tomohiro Tachi (b. Japan)*
*Profession: Associate Professor*
*Fields of interest: Origami, Structural Morphology*
*Address: 3-8-1 Komaba, Meguro-Ku, Tokyo*
*E-mail: tachi@idea.c.u-tokyo.ac.jp*
*Home-page: origami.c.u-tokyo.ac.jp*
*Awards: IASS Tsuboi Award, National Academy of Science Cozzarelli Prize*
*Publications and/or Exhibitions: T. Tachi (2019); Introduction to Structural Origami, Journal of the IASS, 60(1), pp. 7-18*

**Abstract:** *In this presentation, we show geometric properties of a family of polyhedra obtained by folding a regular tetrahedron along regular triangular grids. Each polyhedron is identified by a pair of nonnegative integers. The polyhedron can be cut along a geodesic strip of triangles to be decomposed and unfolded into one or multiple bands (homeomorphic to a cylinder). The number of bands is the greatest common divisor of the two numbers. By a proper choice of pairs of numbers, we can create a common triangular band that folds into different multiple polyhedra that belongs to the family.*

## 1   OVERVIEW

In this paper, we consider a family of polyhedra obtained by folding a regular tetrahedron along a regular triangular grid (Fig. 2). Specifically, for each triangular face, overlay a regular triangular grid on the face such that the vertices of the face lie exactly on the grid points. Now subdividing the polyhedron along such a grid and constraining each edge to be equal lengths will create a new deltahedron. The subdivision can be identified by two nonnegative numbers $a$ and $b$; this is a common notation for creating geodesic polyhedra. Instead of projecting vertices on a sphere like in the geodesic dome, our construction considers making all faces regular triangles, yielding a family of deltahedra. Although we finally consider a folding from the unit tetrahedra, the notation is easier to be understood by placing the unfolding on to the unit triangular grid as in Fig. 3 by appropriate scaling. Then, the numbers $(a, b)$ corresponds to the case when an edge vector is $a\mathbf{e}_x + b\mathbf{e}_y$ when $\mathbf{e}_x$ and $\mathbf{e}_y$ represent the unit vectors along the triangular grid forming $60°$ with respect to each other.

Such deltahedral shapes have been constructed in the context of virus modeling (Caspar and Klug 1962) or appear in the art of modular origami constructions (Kawamura 2015). Gailiunas 2014 computes this family in their work. However, the geometric properties of the family of polyhedra are not yet fully understood. Even its existence, i.e., embedding to Euclidean 3-space, is unknown for a larger $a$ and $b$. We call such a construction "geodesic folding" as the subdivision along the geodesic lines give shapes isometric to the original polyhedron.

In this paper, we focus on the geodesic folding of a regular tetrahedron and show some basic properties of the family. Namely, here are our contributions.
- We show that the deltahedron is decomposable into 1 or multiple triangular bands along the geodesics of the surface, and the number of bands is the greatest common divisor of $a$ and $b$. This formally completes the finding on the mad weaving of tetrahedron by (Gailiunas 2017).
- We show pairs of polyhedra $(a, b)$ and $(a', b')$ that can be folded from the same band.
- We symbolically and numerically construct polyhedra and discuss the volume-increasing nature of the folding.
- We give some open questions regarding its folded states and the volumes.

## 2 GEODESIC BANDS

We claim that deltahedron $(a, b)$ comprises $k$ triangular bands along the geodesics of the surface, where $k$ is the greatest common devisor of $a$ and $b$. This can be shown by considering the tiling of the unfolding of a regular tetrahedron [Akiyama2007] embedded on a regular triangular grid. Now a geodesic strip of triangles on the tetrahedral surface can be drawn as a horizontal straight strip in its unfolding. We would like to know the maximum lengths of such a strip such that it does duplicate the same triangle of the deltahedron. By the property of the tiling of tetrahedral unfolding, the same triangle maps to its $180°$ rotation around the midpoint of the tetrahedral edge, or its translation by $m\mathbf{u} + n\mathbf{v}$ ($m, n \in \mathbb{Z}$), where $\mathbf{u} = (2a, 2b)$ and $\mathbf{v} = (-2b, 2a + 2b)$ represents the translation vectors that form the tiling. A single strip cannot contain $180°$ rotations of the same triangle as the midpoint of corresponding triangles will be a grid point. So, the strip ends exactly when it hits the translational copy of the first triangle; because the strip connects from the opposite side, this part forms a geodesic band of triangles.

Now, the horizontal dimension $(x, 0)$ of the unit strip with $2x$ triangles needs to be represented as $m\mathbf{u} + n\mathbf{v}$. This gives that $2x = \frac{1}{k}S(a, b)$ where $S(a, b) = 4(a^2 + b^2 + ab)$ is the number of triangles on the tetrahedron. This means that the deltahedron is decomposed into $k$ bands of equal number of triangles. In particular, if $a$ and $b$ are coprime, the deltahedron consists of a single band.

## 3  COMMON UNFOLDING

We find a set of pairs of coprime integers, such that each pair unfolds to the geodesic band of the same length. This is possible when for two pairs $(a,b)$ and $(a',b')$, $S(a,b) = S(a',b')$. We found examples of such numbers:

$\{(1,9),(5,6)\} \to 4 \times 91$, $\{(3,40),(8,37),(15,32),(23,25)\} \to 4 \times 1729$, $\{(4,51),(15,44),(19,41),(25,36)\} \to 4 \times 2821$ by enumerating up to $4 \times 5000$-hedra. Such a set of pairs allows a single band to fold into different deltahedra (beyond its mirror image).

## 4  EXTRINSIC PROPERTIES

The extrinsic properties in Euclidean 3-space remain mostly unexplored yet including its existence. Trivially, $(n,0)$: regular tetrahedron, $(1,1)$: five tetrahedra gathered, $(2,1)$: four tetrahedra attached to an icosahedron exist. We found a symbolic form of $(2,2)$ represented by their fold angles as the roots of sextic equations. Other deltahedra with larger numbers were constructed numerically, and their existence is unknown. We implemented the construction of deltahedron through initial guess and dynamic relaxation using *Kangaroo* component of *Grasshopper*.

The length constraint generally gives multiple stable positions with the versions with dimples. We naturally expect that there is a canonical (symmetric) form for each pair $(a,b)$ that maximizes its volume. We also conjecture that the volume maximizes if and only if every vertex pops up, i.e., has an interior solid angle in $(0,2\pi)$. The volumes of the deltahedra are also very interesting, as they increase compared to the regular tetrahedron, but it does not seem to keep increasing as the numbers grow. The table shows the sequence of volumes computed from the numerical method; this suggests that $(5,5)$ may be the one with maximum volume.

## ACKNOWLEDGMENT

This project started in the class "individual and group" organized by Asao Tokolo and Tomohiro Tachi in February 2019. We thank helpful advices from Asao Tokolo and other participants. This work is supported by Japan Society for the Promotion of Science KAKENHI 16H06106.

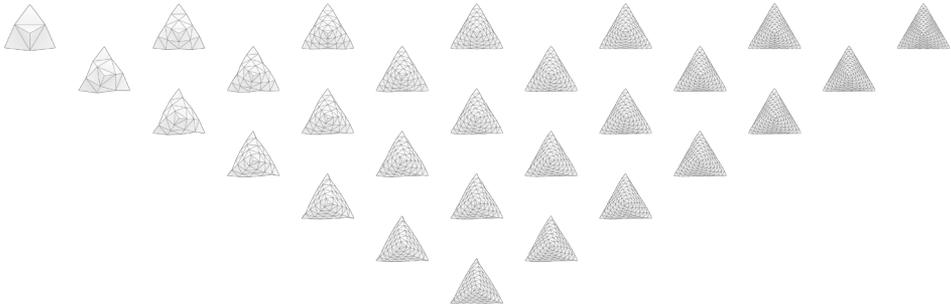

Fig. 1 Numerically computed geodesic folding of regular tetrahedra.

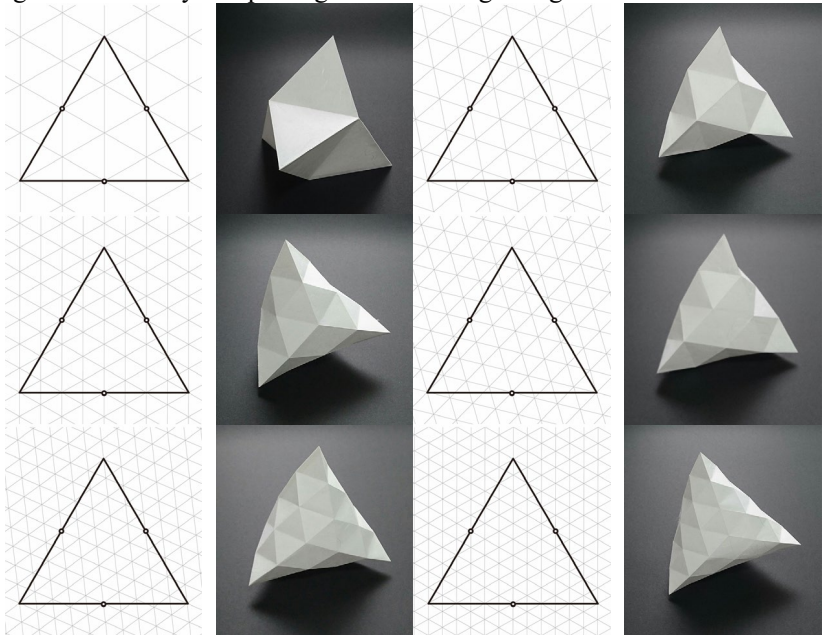

Fig. 2 Geodesic folding of regular tetrahedra corresponding to numbers (1,1), (2,1), (2,2), (3,1), (3,2), (3,3) from top left to bottom right in row major.

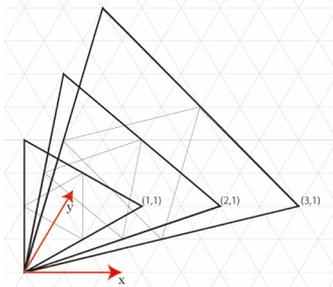

Fig. 3 Unfolding overlaid on a unit grid.

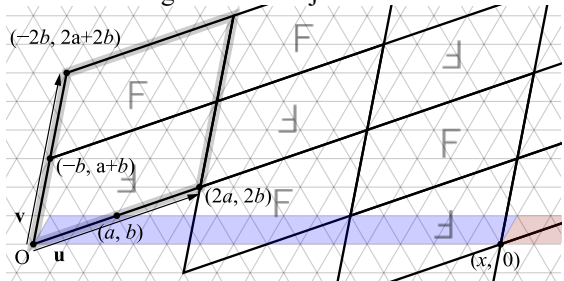

Fig. 4 A horizontal triangular strip on the tiling of the unfolding of tetrahedron.

|   | 1 | 2 | 3 | 4 | 5 | 6 | 7 |
|---|---|---|---|---|---|---|---|
| 1 | 0.96225 | 1.21555 | 1.305946 | 1.311752 | 1.309293 | 1.291614 | 1.27254 |
| 2 |   | 1.29799 | 1.360066 | 1.386255 | 1.388193 | 1.379776 | 1.364827 |
| 3 |   |   | 1.391266 | 1.409836 | 1.415833 | 1.411978 | 1.402804 |
| 4 |   |   |   | 1.422139 | 1.426613 | 1.424967 | 1.41858 |
| 5 |   |   |   |   | **1.43013** | 1.429054 | 1.424419 |
| 6 |   |   |   |   |   | 1.428434 | 1.424969 |
| 7 |   |   |   |   |   |   | 1.4225 |

Table: Volumes of the series of polyhedra relative to unit tetrahedon.